\documentclass[12pt]{article}
\usepackage[totalwidth=460truept,totalheight=610truept]{geometry}
\usepackage{amsfonts,latexsym,amssymb,amsmath,graphicx,accents,slashed,subfigure}
\usepackage[T1]{fontenc}
\usepackage[hidelinks]{hyperref}

\def\theequation{\arabic{section}.\arabic{equation}}

\renewcommand{\theequation}{\thesection.\arabic{equation}}
\global\arraycolsep=1truept

\numberwithin{equation}{section}
\renewcommand{\theequation}{\arabic{section}.\arabic{equation}}

\begin{document}

\bigskip \phantom{C}

\vskip2truecm

\begin{center}
{\huge \textbf{Quantum Gravity, Fakeons}}

\vskip.4truecm

{\huge \textbf{And Microcausality}}

\vskip1truecm

\textsl{Damiano Anselmi\footnote{%
damiano.anselmi@unipi.it} and Marco Piva\footnote{%
marco.piva@df.unipi.it}}

\vskip .1truecm

\textit{Dipartimento di Fisica ``Enrico Fermi'', Universit\`{a} di Pisa, }

\textit{Largo B. Pontecorvo 3, 56127 Pisa, Italy}

\textit{and INFN, Sezione di Pisa,}

\textit{Largo B. Pontecorvo 3, 56127 Pisa, Italy}

\vskip2truecm

\textbf{Abstract}
\end{center}

We investigate the properties of fakeons in quantum gravity at one loop. The
theory is described by a graviton multiplet, which contains the fluctuation $%
h_{\mu \nu }$ of the metric, a massive scalar $\phi $ and the spin-2 fakeon $%
\chi _{\mu \nu }$. The fields $\phi $ and $\chi _{\mu \nu }$ are introduced
explicitly at the level of the Lagrangian by means of standard procedures.
We consider two options, where $\phi $ is quantized as a physical particle
or a fakeon, and compute the absorptive part of the self-energy of the
graviton multiplet. The width of $\chi _{\mu \nu }$, which is negative,
shows that the theory predicts the violation of causality at energies larger
than the fakeon mass. We address this issue and compare the results with
those of the Stelle theory, where $\chi _{\mu \nu }$ is a ghost instead of a
fakeon.

\vfill\eject

\section{Introduction}

\label{intro} \setcounter{equation}{0}

A theory of quantum gravity was formulated in ref. \cite{LWgrav} by means of
a new prescription to treat the poles of the free propagators and turn the
ghosts due to the higher derivatives into \textit{fakeons} \cite{fakeons}.
The classical Lagrangian contains the Hilbert term, the quadratic terms $%
\sqrt{-g}R_{\mu \nu }R^{\mu \nu }$ and $\sqrt{-g}R^{2}$ and the cosmological
term. The fakeons are \textquotedblleft fake particles\textquotedblright ,
which contribute to the correlation functions, but disappear from the
physical spectrum. The idea takes inspiration from the Lee-Wick models \cite%
{leewick,LW}, in particular their reformulation as nonanalytically Wick
rotated Euclidean theories \cite{LWformulation,LWunitarity}. An essentially
unique\footnote{%
This means that the action has a finite number of independent parameters and
admits a finite number (two, in our case) of physically consistent
quantization prescriptions.} strictly renormalizable theory of quantum
gravity emerges from this approach, which is perturbatively unitary up to
the effects due to the cosmological constant\footnote{%
A consistent theory of scattering with the properties we need may not exist
at nonvanishing cosmological constant. On this topic, see the discussions of
refs. \cite{adsscatt}. The problem concerns every realistic theory of
quantum gravity, including the low-energy nonrenormalizable one, which can
be used as an effective field theory.}.

In this paper, we investigate the properties of the fakeons in quantum
gravity at one loop. To begin with, we introduce auxiliary fields and make
changes of field variables, to finalize a number of arguments that are
available in the literature \cite{stuck} and convert the higher-derivative
action of \cite{LWgrav} into an equivalent action that does not contain
higher derivatives and is organized so as to fully diagonalize the kinetic
part in the nonlinear case. The new setting is convenient to calculate the
quantities we are interested in here. It is not equally convenient to study
the renormalization of the theory (which is not affected by the fakeon
prescription and has been already studied in a variety of approaches \cite%
{tonin,barvinsky,percacci,agravity,agravity2,UVQG}).

Quantum gravity is described by a \textit{graviton multiplet}, made of the
fluctuation $h_{\mu \nu }$ of the metric tensor around flat space, a massive
scalar $\phi $ and a massive spin-2 field $\chi _{\mu \nu }$. To have
perturbative unitarity (up to the effects of the cosmological constant) the
field $\chi _{\mu \nu }$ must be quantized as a fakeon, because its
quadratic action carries the wrong overall sign. Instead, the quadratic
action of $\phi $ carries the right overall sign, so $\phi $ can be
quantized either as a fakeon or a physical particle. This leads to two
possibilities, which we call \textit{graviton/fakeon/fakeon (GFF) theory}
and \textit{graviton/scalar/fakeon (GSF) theory}, respectively.

We study the absorptive part of the self-energy of the graviton multiplet in
both cases. A number of techniques to calculate this quantity and, more
generally, deal with the fakeons, have been developed in ref. \cite{UVQG}.
The approach we follow here further simplifies the computations and allows
us to extend the results in several directions. In particular, we obtain the
width $\Gamma _{\chi }$ of the spin-2 fakeon $\chi _{\mu \nu }$, which is
related to the central charge $C$ of the matter fields, and the width $%
\Gamma _{\phi }$ of $\phi $. The value of $\Gamma _{\chi }$ is negative,
which means that $\chi _{\mu \nu }$ is responsible for the violation of
microcausality. At center-of-mass energies close to the fakeon mass $m_{\chi
}$, and for time intervals of the order of $1/|\Gamma _{\chi }| $ (referred
to the center-of-mass frame) the common notions of past, present and future,
as well as cause and effect, lose meaning. Two events can be related in a
causal way only if they are separated by a time interval that is much longer
than $1/|\Gamma _{\chi }|$.

The breakdown of causality at very small distances is expected, because it
is also a property of the Lee-Wick models, where it has been studied in
detail \cite{leewick,LW,coleman}. Although the quantum gravity theory of 
\cite{LWgrav} is not of the Lee-Wick type, the fakeon quantization
prescription introduces an infinitesimal width that turns the theory into a
Lee-Wick model in an intermediate step. From the physical point of view, we
do not have arguments to claim that nature must be causal up to infinite
energies, so we regard the violation of microcausality as a key prediction
of quantum gravity.

We also compare the results of the GFF and GSF theories with those of the
Stelle\ theory \cite{stelle}, recently considered by Salvio and Strumia from
the phenomenological point of view in refs. \cite{agravity,agravity2}, which
is a \textit{graviton/scalar/ghost} (GSGh) theory. The classical action of
the GSGh theory is the same, but its quantization is different in that the
Feynman prescription is used for all the poles of the free propagators,
including the one of $\chi _{\mu \nu }$. Then $\chi _{\mu \nu }$ is a ghost,
instead of a fakeon, and does contribute to the absorptive parts, as well as
the central charge $C$. The quantities we calculate do not exhibit important
differences up to energies equal to the fakeon mass $m_{\chi }$. For
example, the width $\Gamma _{\chi }$ is the same in the GSF and GSGh
theories. The differences start to become important above $m_{\chi }$, where
the optical theorem is violated in the GSGh theory.

The computations are performed at vanishing cosmological constant $\Lambda
_{C}$, since the corrections due to $\Lambda _{C}$ are too small for the
quantities we study. The results of ref. \cite{UVQG} are recovered as a
particular case. We include results for Proca vectors and Pauli-Fierz spin-2
fields.

The paper is organized as follows. In section \ref{fakeon} we isolate the
fakeons by working out an equivalent action of quantum gravity that does not
contain higher derivatives. In section \ref{quantization} we outline the
prescriptions to quantize the theory. In section \ref{abso} we calculate the
absorptive part of the self-energy of the graviton multiplet. In section \ref%
{probe} we calculate the width $\Gamma _{\chi }$ of $\chi _{\mu \nu }$ and
discuss the relation between $\Gamma _{\chi }$ and the central charge $C$,
as well as the violations of microcausality. We also give the width of $\phi 
$. In section \ref{SScompa} we extend the calculations to the Stelle theory
and compare the results with those of the GFF\ and GSF\ theories. Section %
\ref{concl} contains the conclusions. The appendices \ref{appA} and \ref%
{appB} contain details about some tools used for the calculations and other
results about the absorptive parts.

\section{Isolating the fakeons in quantum gravity}

\setcounter{equation}{0} \label{fakeon}

The theory of quantum gravity (coupled to matter) proposed in ref. \cite%
{LWgrav} has action%
\begin{equation}
S_{\text{QG}}=-\frac{1}{2\kappa ^{2}}\int \sqrt{-g}\left[ 2\Lambda
_{C}+\zeta R+\alpha \left( R_{\mu \nu }R^{\mu \nu }-\frac{1}{3}R^{2}\right) -%
\frac{\xi }{6}R^{2}\right] +S_{m}(g,\Phi ),  \label{SQG}
\end{equation}%
where $\alpha $, $\xi $, $\zeta $, $\Lambda _{C}$ and $\kappa $ are real
constants, with $\alpha >0$, $\xi >0$ and $\zeta >0$, and $S_{m}$ is the
action of the matter sector. For example, we can take $S_{m}$ as the
covariantized action of the standard model, or one of its popular
extensions, equipped with the nonminimal couplings that are compatible with
the renormalizability.

In this section we isolate the fakeons by means of auxiliary fields and
field redefinitions. We obtain an equivalent action that does not contain
higher-derivatives and is useful for the calculations of the next sections.
In particular, we fully diagonalize the kinetic part in the nonlinear case.
In the next section we explain how to quantize the theory in the new setting.

To our knowledge, the new action, which is given by formula (\ref{sew}), is
not available in the literature in a complete form. Partial derivations can
nevertheless be found. For example, the authors of \cite{stuck} work at $%
\Lambda _{C}=0$, with no matter sector $S_{m}$ and stop short of finalizing
the action to concentrate on the analysis of the quadratic part around flat
space, since their main interest is to highlight the degrees of freedom.

We assume that $S_{m}$ is at least quadratic in the matter fields $\Phi $.
For simplicity, we work with bosonic fields. The arguments can be easily
generalized to fermionic fields by using the tetrad formalism.

Defining%
\begin{equation*}
\hat{\Lambda}_{C}=\Lambda _{C}\left( 1+\frac{4}{3}\frac{\xi \Lambda _{C}}{%
\zeta ^{2}}\right) ,\qquad \hat{\zeta}=\zeta \frac{\hat{\Lambda}_{C}}{%
\Lambda _{C}},\qquad \hat{R}_{\mu \nu }=R_{\mu \nu }+\frac{\Lambda _{C}}{%
\zeta }g_{\mu \nu },\qquad \hat{R}=R+\frac{4\Lambda _{C}}{\zeta },
\end{equation*}%
and adding the integral of a total derivative, the action (\ref{SQG}) can be
written in the more convenient form%
\begin{equation*}
S_{\text{QG}}=\hat{S}_{\text{HE}}(g)+S_{\text{W}}(g)+\frac{\xi }{12\kappa
^{2}}\int \sqrt{-g}\hat{R}^{2}+S_{m}(g,\Phi ),
\end{equation*}%
where 
\begin{equation}
\hat{S}_{\text{HE}}(g)=-\frac{1}{2\kappa ^{2}}\int \sqrt{-g}\left( 2\hat{%
\Lambda}_{C}+\hat{\zeta}R\right)   \label{SHE}
\end{equation}%
is the Hilbert-Einstein action and%
\begin{equation*}
S_{\text{W}}(g)=-\frac{\alpha }{4\kappa ^{2}}\int \sqrt{-g}C_{\mu \nu \rho
\sigma }C^{\mu \nu \rho \sigma }
\end{equation*}%
is the Weyl action, $C_{\mu \nu \rho \sigma }$ denoting the Weyl tensor.

\subsection{Step 1: massive scalar}

We introduce an auxiliary field $\hat{\phi}$ and write $S_{\text{QG}}$ as%
\begin{equation*}
S_{\text{QG}}=\hat{S}_{\text{HE}}(g)+S_{\text{W}}(g)+\frac{\xi }{12\kappa
^{2}}\int \sqrt{-g}(2\hat{R}-\hat{\phi})\hat{\phi}+S_{m}(g,\Phi ).
\end{equation*}%
Then we perform the Weyl transformation 
\begin{equation}
g_{\mu \nu }\rightarrow g_{\mu \nu }\mathrm{e}^{\kappa \phi },  \label{weyl}
\end{equation}%
where%
\begin{equation}
\phi =-\frac{1}{\kappa }\ln \left( 1-\frac{\xi \hat{\phi}}{3\hat{\zeta}}%
\right) .  \label{phitilde}
\end{equation}%
So doing, we obtain the equivalent action%
\begin{equation}
S_{\text{QG}}=\hat{S}_{\text{HE}}(g)+S_{\text{W}}(g)+S_{\phi }(g,\phi
)+S_{m}(g\mathrm{e}^{\kappa \phi },\Phi ),  \label{sqgeq}
\end{equation}%
where 
\begin{equation}
S_{\phi }(g,\phi )=\frac{3\hat{\zeta}}{4}\int \sqrt{-g}\left[ D_{\mu }\phi
D^{\mu }\phi -\frac{m_{\phi }^{2}}{\kappa ^{2}}\left( 1-\mathrm{e}^{\kappa
\phi }\right) ^{2}\right] ,  \label{sphi}
\end{equation}%
the squared mass of $\phi $ being 
\begin{equation}
m_{\phi }^{2}=\frac{\zeta }{\xi }.  \label{mphi}
\end{equation}

\subsection{Step 2: spin-2 fakeon}

Now we take care of the spin-2 fakeon. We have 
\begin{equation*}
\hat{S}_{\text{HE}}(g)+S_{\text{W}}(g)=\tilde{S}_{\text{HE}}(g)-\frac{\alpha 
}{2\kappa ^{2}}\int \sqrt{-g}\left( \tilde{R}_{\mu \nu }\tilde{R}^{\mu \nu }-%
\frac{1}{3}\tilde{R}^{2}\right) ,
\end{equation*}%
up to the integral of a total derivative, where%
\begin{eqnarray*}
\tilde{S}_{\text{HE}}(g) &=&-\frac{1}{2\kappa ^{2}}\int \sqrt{-g}\left( 2%
\tilde{\Lambda}_{C}+\tilde{\zeta}R\right) ,\qquad \tilde{R}_{\mu \nu }=\hat{R%
}_{\mu \nu }, \\
\tilde{\Lambda}_{C} &=&\hat{\Lambda}_{C}\left( 1+\frac{2}{3}\frac{\alpha 
\hat{\Lambda}_{C}}{\hat{\zeta}^{2}}\right) =\Lambda _{C}\left( 1+\frac{2}{3}%
\frac{(\alpha +2\xi )\Lambda _{C}}{\zeta ^{2}}\right) ,\qquad \tilde{\zeta}=%
\hat{\zeta}\frac{\tilde{\Lambda}_{C}}{\hat{\Lambda}_{C}}=\zeta \frac{\tilde{%
\Lambda}_{C}}{\Lambda _{C}}.
\end{eqnarray*}

We introduce auxiliary fields $\chi _{\mu \nu }$ by writing the action $S_{%
\text{QG}}$ as%
\begin{eqnarray}
S_{\text{QG}} &=&\tilde{S}_{\text{HE}}(g)-\frac{\tilde{\zeta}}{2\kappa ^{2}}%
\int \sqrt{-g}\left[ 2\chi ^{\mu \nu }\left( \tilde{R}_{\mu \nu }-\frac{1}{2}%
g_{\mu \nu }\tilde{R}\right) -\frac{\tilde{\zeta}}{\alpha }(\chi _{\mu \nu
}\chi ^{\mu \nu }-\chi ^{2})\right]   \notag \\
&&+S_{\phi }(g,\phi )+S_{m}(g\mathrm{e}^{\kappa \phi },\Phi ),  \label{sqg}
\end{eqnarray}%
where $\chi =\chi _{\mu \nu }g^{\mu \nu }$. At this point, we perform the
metric-tensor redefinition 
\begin{equation}
g_{\mu \nu }\rightarrow g_{\mu \nu }+2\chi _{\mu \nu }+\chi _{\mu \nu }\chi
-2\chi _{\mu \rho }\chi _{\nu }^{\rho }\equiv g_{\mu \nu }+\psi _{\mu \nu }.
\label{metricredef}
\end{equation}%
The linear contribution to $\psi _{\mu \nu }$ is fixed so that the
transformed action contains no terms that are linear in $\chi _{\mu \nu }$.
The quadratic corrections are determined so that the mass terms of the $\chi
_{\mu \nu }$ action get the right Pauli-Fierz form and the limit $\Lambda
_{C}\rightarrow 0$ remains regular.

Applying the redefinition (\ref{metricredef}) to (\ref{sqg}), we obtain the
equivalent action of quantum gravity we are going to work with in this
paper, which reads%
\begin{equation}
S_{\text{QG}}(g,\phi ,\chi ,\Phi )=\tilde{S}_{\text{HE}}(g)+S_{\chi }(g,\chi
)+S_{\phi }(g+\psi ,\phi )+S_{m}(g\mathrm{e}^{\kappa \phi }+\psi \mathrm{e}%
^{\kappa \phi },\Phi ),  \label{sew}
\end{equation}%
where%
\begin{equation}
S_{\chi }(g,\chi )=\tilde{S}_{\text{HE}}(g+\psi )-\tilde{S}_{\text{HE}%
}(g)+\int \left[ -2\chi _{\mu \nu }\frac{\delta \tilde{S}_{\text{HE}}(g)}{%
\delta g_{\mu \nu }}+\frac{\tilde{\zeta}^{2}}{2\alpha \kappa ^{2}}\sqrt{-g}%
(\chi _{\mu \nu }\chi ^{\mu \nu }-\chi ^{2})\right] _{g\rightarrow g+\psi }
\label{spsi}
\end{equation}%
is the action of the fakeon $\chi _{\mu \nu }$. We find%
\begin{equation}
S_{\chi }(g,\chi )=-\frac{\tilde{\zeta}}{\kappa ^{2}}S_{\text{PF}}(g,\chi
,m_{\chi }^{2})-\frac{\tilde{\zeta}}{2\kappa ^{2}}\int \sqrt{-g}R^{\mu \nu
}(\chi \chi _{\mu \nu }-2\chi _{\mu \rho }\chi _{\nu }^{\rho })+S_{\chi
}^{(>2)}(g,\chi ),  \label{scc}
\end{equation}%
where%
\begin{eqnarray}
S_{\text{PF}}(g,\chi ,m_{\chi }^{2}) &=&\frac{1}{2}\int \sqrt{-g}\left[
D_{\rho }\chi _{\mu \nu }D^{\rho }\chi ^{\mu \nu }-D_{\rho }\chi D^{\rho
}\chi +2D_{\mu }\chi ^{\mu \nu }D_{\nu }\chi -2D_{\mu }\chi ^{\rho \nu
}D_{\rho }\chi _{\nu }^{\mu }\right.  \notag \\
&&\left. -m_{\chi }^{2}(\chi _{\mu \nu }\chi ^{\mu \nu }-\chi ^{2})\right]
\label{SPF}
\end{eqnarray}%
is the covariantized Pauli-Fierz action and $S_{\chi }^{(>2)}(g,\chi )$ are
corrections that are at least cubic in $\chi $. The squared mass of the
spin-2 fakeon is%
\begin{equation}
m_{\chi }^{2}=\frac{\tilde{\zeta}}{\alpha }.  \label{mchi}
\end{equation}

The transformations (\ref{weyl}), (\ref{phitilde}) and (\ref{metricredef})
are ultralocal (i.e. they depend on the fields, but not their derivatives),
so the Jacobians are identically one in dimensional regularization. This
means that we can use the new action $S_{\text{QG}}(g,\phi ,\chi ,\Phi )$ of
formula (\ref{sew}) as the action of quantum gravity at the level of the
functional integral.

So far, we have kept the cosmological constant different from zero, but in
many situations it may be neglected. When that is the case, it is convenient
to replace the field redefinition (\ref{metricredef}) with%
\begin{equation}
g_{\mu \nu }\rightarrow g_{\mu \nu }+2\chi _{\mu \nu },  \label{metricredef2}
\end{equation}%
so that, instead of (\ref{sew}),\ we have%
\begin{equation}
S_{\text{QG}}(g,\phi ,\chi ,\Phi )=S_{\text{H}}(g)+S_{\chi }^{\prime
}(g,\chi )+S_{\phi }(g+2\chi ,\phi )+S_{m}(g\mathrm{e}^{\kappa \phi }+2\chi 
\mathrm{e}^{\kappa \phi },\Phi ),  \label{nocosmo}
\end{equation}%
where%
\begin{equation*}
S_{\text{H}}(g)=-\frac{\zeta }{2\kappa ^{2}}\int \sqrt{-g}R,
\end{equation*}%
is the Hilbert action and $S_{\chi }^{\prime }(g,\chi )$ is the new $\chi $
action, still given by (\ref{spsi}), but with $\Lambda _{C}=0$ and $\psi
_{\mu \nu }$ replaced by $2\chi _{\mu \nu }$. We find%
\begin{eqnarray}
S_{\chi }^{\prime }(g,\chi ) &=&-2\int \frac{\delta ^{2}S_{\text{H}}}{\delta
g_{\mu \nu }(x)\delta g_{\rho \sigma }(y)}\chi _{\mu \nu }(x)\chi _{\rho
\sigma }(y)\mathrm{d}x\mathrm{d}y+\frac{\zeta ^{2}}{2\alpha \kappa ^{2}}\int 
\sqrt{-g}(\chi _{\mu \nu }\chi ^{\mu \nu }-\chi ^{2})  \notag \\
&&-\frac{8}{3}\int \frac{\delta ^{3}S_{\text{H}}}{\delta g_{\mu \nu
}(x)\delta g_{\rho \sigma }(y)\delta g_{\alpha \beta }(z)}\chi _{\mu \nu
}(x)\chi _{\rho \sigma }(y)\chi _{\alpha \beta }(z)\mathrm{d}x\mathrm{d}y%
\mathrm{d}z  \label{schi} \\
&&+\frac{\zeta ^{2}}{2\alpha \kappa ^{2}}\int \sqrt{-g}(5\chi \chi _{\mu \nu
}\chi ^{\mu \nu }-4\chi _{\mu \nu }\chi ^{\mu \rho }\chi _{\rho }^{\nu
}-\chi ^{3})+S_{\chi }^{(>3)}(g,\chi ).  \notag
\end{eqnarray}%
where $S_{\chi }^{(>3)}(g,\chi )$ are corrections that are at least quartic
in $\chi _{\mu \nu }$, which are not needed in the calculations of this
paper. Note that the nonminimal couplings of the quadratic part 
\begin{equation*}
-\frac{\zeta }{\kappa ^{2}}S_{\text{PF}}(g,\chi ,m_{\chi }^{2})-\frac{\zeta 
}{4\kappa ^{2}}\int \sqrt{-g}\left( 4\chi \chi _{\mu \nu }R^{\mu \nu }-8\chi
_{\mu \nu }\chi ^{\nu \rho }R_{\rho }^{\mu }+2R\chi _{\mu \nu }\chi ^{\mu
\nu }-R\chi ^{2}\right)
\end{equation*}%
of $S_{\chi }^{\prime }(g,\chi )$ differ from those of (\ref{scc}), and the $%
\chi $ squared mass is now%
\begin{equation}
m_{\chi }^{2}=\frac{\zeta }{\alpha }.  \label{mchi2}
\end{equation}%
Formulas (\ref{spsi}) and (\ref{schi}) show that the vertices of the $\chi $
actions are related to the vertices of the Hilbert-Einstein action, apart
from corrections proportional to $m_{\chi }^{2}$.

The new actions (\ref{sew}) and (\ref{nocosmo}) are convenient to calculate
the quantities we are interested in, but make the renormalizability of the
theory much less evident than it was in the original field variables (\ref%
{SQG}). On general grounds, the only effect of a perturbative change of
field variables on the divergent sector of the theory is to require extra
field renormalizations, which are generically nonpolynomial, yet
perturbatively local. A precise match between the divergent parts,
calculated before and after the field redefinition, can be worked out by
relating them to the renormalizations of the composite operators involved in
the transformation \cite{fieldcov}.

\section{Quantization}

\setcounter{equation}{0} \label{quantization}

Expanding the metric tensor around flat space as $g_{\mu \nu }=\eta _{\mu
\nu }+2\kappa h_{\mu \nu }$, where $\eta _{\mu \nu }=$ diag$(1,$ $-1,$ $-1,$ 
$-1)$, the graviton sector is described by the \textit{graviton multiplet}%
\begin{equation}
G_{A}=\{h_{\mu \nu },\phi ,\chi _{\rho \sigma }\},  \label{gmulti}
\end{equation}%
made of the fluctuation $h_{\mu \nu }$ of the metric, the massive scalar $%
\phi $ and the massive spin-2 field $\chi _{\mu \nu }$.

Assuming that $|\Lambda _{C}|$ is sufficiently small, so that both $\tilde{%
\zeta}$ and $\hat{\zeta}$ are positive, the action $S_{\chi }$ of formula (%
\ref{scc})\ carries the wrong overall sign. This means that, to have
perturbative unitarity (up to corrections due to the cosmological constant), 
$\chi _{\mu \nu }$ must be quantized as a fakeon, following the prescription
of ref. \cite{LWgrav}. Instead, the quadratic action $S_{\phi }$ of eq. (\ref%
{sphi}) carries the right overall sign, so $\phi $ can be quantized either
as a fakeon or a physical particle. This leads to two possibilities, which
we call \textit{graviton/fakeon/fakeon (GFF) theory} and \textit{%
graviton/scalar/fakeon (GSF) theory}, respectively. Being perturbatively
unitary (up to the effects of the cosmological constant) and renormalizable,
they are both good candidates to describe quantum gravity. We could also
view $\phi $ and $\chi _{\mu \nu }$ as part of the matter sector.

We define the GFF\ and GSF\textit{\ }prescriptions by introducing two
infinitesimal widths $\epsilon $ and $\mathcal{E}$ in the propagators\ as
follows:

($a$) replace $p^{2}$ with $p^{2}+i\epsilon $ everywhere in the denominators
of the propagators, where $p$ denotes the momentum;

($b$) turn the $\chi $ poles into fakeons by means of the replacement%
\begin{equation}
\frac{1}{p^{2}-m_{\chi }^{2}+i\epsilon }\rightarrow \frac{p^{2}-m_{\chi }^{2}%
}{(p^{2}-m_{\chi }^{2}+i\epsilon )^{2}+\mathcal{E}^{4}};  \label{nopri}
\end{equation}

($c$) [only in the GFF case] turn the $\phi $ poles into fakeons by means of
the replacement 
\begin{equation}
\frac{1}{p^{2}-m_{\phi }^{2}+i\epsilon }\rightarrow \frac{p^{2}-m_{\phi }^{2}%
}{(p^{2}-m_{\phi }^{2}+i\epsilon )^{2}+\mathcal{E}^{4}}.  \label{nopri2}
\end{equation}

($d$) calculate the diagrams in the Euclidean framework, nonanalytically
Wick rotate them as explained in refs. \cite%
{LWformulation,LWunitarity,fakeons}, then make $\epsilon $ tend to zero
first and $\mathcal{E}$ tend to zero last.

Note that because of the Wick rotation involved in point ($d$) the
distributions appearing on the right-hand sides of eqs. (\ref{nopri}) and (%
\ref{nopri2}) do not give the principal value (which would require to
integrate on real energies).

An equivalent, and often more efficient, way to formulate the
graviton/fakeon prescription is to combine point ($a$) with the requirement
that, in evaluating the loop integrals,

($a^{\prime }$) every threshold involving a fakeon must be overcome by means
of the \textit{average continuation}, which is the arithmetic average of the
two analytic continuations that circumvent the threshold.

The space of the complexified external momenta is divided into disjoint
regions of analyticity. All of them can be unambiguously reached from the
Euclidean region by means of the average continuation.

The free propagator of the metric fluctuation $h_{\mu \nu }$ reads%
\begin{equation}
\langle h_{\mu \nu }(p)\hspace{0.01in}\hspace{0.01in}h_{\rho \sigma
}(-p)\rangle _{0}=\frac{i\mathcal{(}\eta _{\mu \rho }\eta _{\nu \sigma
}+\eta _{\mu \sigma }\eta _{\nu \rho }-\eta _{\mu \nu }\eta _{\rho \sigma })%
}{2\tilde{\zeta}(p^{2}-m_{h}^{2}+i\epsilon )},  \label{propo}
\end{equation}%
where $m_{h}^{2}=-2\Lambda _{C}/\zeta $, in the de Donder gauge. We recall
that the cutting equations \cite{cuttingeq} (which are the diagrammatic
equations that lead to the optical theorem) are formally satisfied even when
cosmological constant is nonvanishing, as long as it is negative \cite%
{LWgrav}, although a consistent theory of scattering likely does not exist
in that case.

The free propagator of $\chi _{\mu \nu }$ reads%
\begin{equation}
\langle \chi _{\mu \nu }(p)\hspace{0.01in}\hspace{0.01in}\chi _{\rho \sigma
}(-p)\rangle _{0}=-\frac{i\kappa ^{2}}{\tilde{\zeta}}\frac{p^{2}-m_{\chi
}^{2}}{(p^{2}-m_{\chi }^{2}+i\epsilon )^{2}+\mathcal{E}^{4}}%
\mbox{{\LARGE
$\Pi $}}_{\mu \nu \rho \sigma }^{(2)}(p,m_{\chi }^{2}),  \label{propchi}
\end{equation}%
where 
\begin{equation}
\mbox{{\LARGE $\Pi $}}_{\mu \nu \rho \sigma }^{(2)}(p,m_{\chi }^{2})=\frac{1%
}{2}\left( \pi _{\mu \rho }\pi _{\nu \sigma }+\pi _{\mu \sigma }\pi _{\nu
\rho }-\frac{2}{3}\pi _{\mu \nu }\pi _{\rho \sigma }\right) ,\qquad \pi
_{\mu \nu }=\eta _{\mu \nu }-\frac{p_{\mu }p_{\nu }}{m_{\chi }^{2}},
\label{proje}
\end{equation}%
are spin-2 and spin-1 on-shell projectors, respectively.

The free $\phi $ propagator reads%
\begin{equation*}
\langle \phi (p)\hspace{0.01in}\hspace{0.01in}\phi (-p)\rangle _{0\text{%
\hspace{0.01in}GFF}}=\frac{2i}{3\hat{\zeta}}\frac{p^{2}-m_{\phi }^{2}}{%
(p^{2}-m_{\phi }^{2}+i\epsilon )^{2}+\mathcal{E}^{4}},\quad \langle \phi (p)%
\hspace{0.01in}\hspace{0.01in}\phi (-p)\rangle _{0\text{\hspace{0.01in}GSF}}=%
\frac{2i}{3\hat{\zeta}}\frac{1}{p^{2}-m_{\phi }^{2}+i\epsilon },
\end{equation*}%
in the GFF and GSF cases, respectively.

The physical fields are the physical components of $h_{\mu \nu }$ (obtained
by projecting away the unphysical components and the Faddeev-Popov ghosts in
the usual ways), the massive scalar $\phi $ (in the GSF theory only) and the
matter fields $\Phi $.

The Fock space $V$ of the physical states is the Hilbert space built as
follows. Consider the states $|n\rangle $ obtained by acting on the vacuum $%
|0\rangle $ by means of the creation operators of the physical fields. Then,
build the metric space $\mathcal{F}$ made of the finite linear combinations
of the states $|n\rangle $. Finally, complete $\mathcal{F}$ to the Hilbert
space $V$ by means of the Cauchy procedure.

The space $V$ is a proper subspace of the total Fock space $W$, which also
contains the states built with the creation operators of the fakeons ($%
a_{\chi }^{\dag }$ in the GSF\ theory and $a_{\chi }^{\dag }$, $a_{\phi
}^{\dag }$ in the GFF theory). The free Hamiltonian $H_{\text{free}}$ is
bounded from below in $V$, although it is not bounded from below in $W$ (due
to the negative contributions brought by $\chi _{\mu \nu }$). Perturbative
unitarity is the statement that the projection from $W$ onto $V$ is
consistent, i.e. the states that are projected away are not generated back
in the cutting equations and the optical theorem. More details are given in
sections \ref{probe} and \ref{SScompa}.

Before turning to the computations, let us recall that the standard
quantization prescription \cite{stelle} is just made of point ($a$) for
every pole. Then $\phi $ is a physical particle, but $\chi _{\mu \nu }$ is a
ghost, due to the overall minus sign that multiplies the right-hand side of (%
\ref{propchi}). In that case the Fock space is the whole $W$.

Another interesting possibility has been pointed out by Avramidi and
Barvinsky in ref. \cite{barvinsky}, where it was noted that for $\Lambda
_{C}>0$, $\xi <0$ the action (\ref{SQG}) is positive definite in the
Euclidean framework and the theory is asymptotically free (when matter is
switched off). However, $\xi <0$ makes the squared mass of $\phi $ negative.
The fakeon prescription of ref. \cite{LWgrav} works for poles located on the
real axis, irrespectively of the sign of the residue at the pole. Tachyons
do not fall in that class, so we cannot guarantee in this moment that a
proper generalization of the prescription (\ref{nopri2}) exists for $\xi <0$.

\section{Absorptive part of the self-energy}

\setcounter{equation}{0} \label{abso}

\begin{figure}[t]
\begin{center}
\includegraphics[width=12truecm]{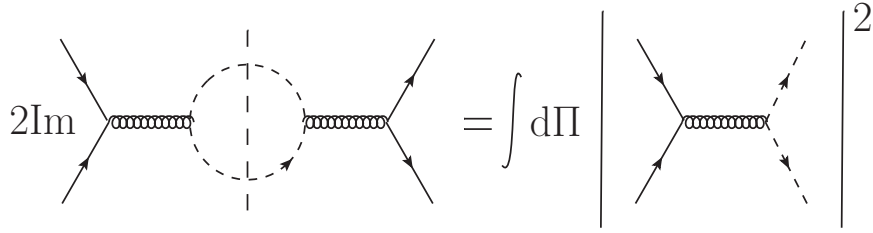}
\end{center}
\caption{Processes involving the absorptive part of the graviton-multiplet
self-energy}
\label{Fig1}
\end{figure}

The absorptive part of the self-energy of the graviton multiplet is
important because it allows us to extract physically observable quantities,
as explained in section \ref{probe}. In Fig. \ref{Fig1} we show a basic
process where the absorptive part plays a key role. On the right-hand side,
we have the squared modulus of the transition amplitude between some initial
states, denoted by the continuous lines, and some final states, denoted by
the dashed lines. The wiggled line denotes the graviton multiplet.
Integrating on the phase space $\Pi $ of the final states, we obtain twice
the imaginary part of the amplitude shown on the left-hand side. In this
section, we ignore the initial states, which leads us to consider the
absorptive parts 
\begin{equation}
M_{AB}\equiv \langle G_{A}(p)G_{B}(-p)\rangle _{\text{abs}}^{\text{1 loop}}
\label{mab}
\end{equation}%
of the matrix $\langle G_{A}G_{B}\rangle $ of the graviton-multiplet
two-point functions at one loop. For simplicity, we set the cosmological
constant to zero, but the procedure can be easily generalized to $\Lambda
_{C}\neq 0$. The gauge-dependent contributions are calculated in the de
Donder gauge. 

We can throw away the diagrams where the fakeons propagate inside the loop.
Indeed, according to the prescription of the previous section, in those
cases we are lead to calculate the average continuation above the
thresholds, which has no absorptive part. Then we can drop $S_{\chi
}^{\prime }(g,\chi )$ from the action (\ref{nocosmo}) and work with the
simplified action%
\begin{equation}
S_{\text{QG}}^{\prime }(g,\phi ,\chi ,\Phi )=S_{\text{H}}(g)+S_{\phi
}(g+2\chi ,\phi )+S_{m}(g\mathrm{e}^{\kappa \phi }+2\chi \mathrm{e}^{\kappa
\phi },\Phi ).  \label{spqg}
\end{equation}%
The tadpole diagrams do not contribute to the absorptive parts, so we can
focus on the cubic vertices. Expanding (\ref{spqg}) to the cubic order in $%
\chi _{\mu \nu }$-$\phi $-$\Phi $, we obtain a further simplified action for
the GSF theory, which is 
\begin{eqnarray}
S_{\text{QG}}^{\text{GSF}}(g,\phi ,\chi ,\Phi ) &=&S_{\text{H}}(g)+S_{\phi
}(g,\phi )+S_{m}(g,\Phi )  \notag \\
&&-\frac{1}{2}\int \sqrt{-g}\left[ (2\chi _{\mu \nu }+\kappa \phi g_{\mu \nu
})T_{m}^{\mu \nu }(g,\Phi )+2\chi _{\mu \nu }T_{\phi }^{\mu \nu }(g,\phi )%
\right] ,  \label{sgsf}
\end{eqnarray}%
where%
\begin{equation}
T_{m}^{\mu \nu }(g,\Phi )=-\frac{2}{\sqrt{-g}}\frac{\delta S_{m}(g,\Phi )}{%
\delta g_{\mu \nu }},\qquad T_{\phi }^{\mu \nu }(g,\phi )=-\frac{2}{\sqrt{-g}%
}\frac{\delta S_{\phi }(g,\phi )}{\delta g_{\mu \nu }},  \label{stress}
\end{equation}%
are the energy-momentum tensor of the matter fields and the one of $\phi $,
respectively.

In the GFF\ case, the field $\phi $ can also be ignored inside the loop,
which means that we can work with%
\begin{equation}
S_{\text{QG}}^{\text{GFF}}(g,\phi ,\chi ,\Phi )=S_{\text{H}}(g)+S_{m}(g,\Phi
)-\frac{1}{2}\int \sqrt{-g}(2\chi _{\mu \nu }+g_{\mu \nu }\kappa \phi
)T_{m}^{\mu \nu }(g,\Phi ).  \label{sgff}
\end{equation}

We collect the results about $M_{AB}$ into the one-loop absorptive part $%
\Gamma _{\text{abs}}$ of the $\Gamma $ functional. We can decompose $\Gamma
_{\text{abs}}$ as 
\begin{eqnarray}
\Gamma _{\text{abs}}^{\text{GFF}} &=&\Gamma _{\text{abs}}^{hh}+\Gamma _{%
\text{abs}}^{m},  \label{gabsGFF} \\
\Gamma _{\text{abs}}^{\text{GSF}} &=&\Gamma _{\text{abs}}^{hh}+\Gamma _{%
\text{abs}}^{\phi h}+\Gamma _{\text{abs}}^{\phi \phi }+\Gamma _{\text{abs}%
}^{m},  \label{gabsGSF}
\end{eqnarray}%
in the GFF and GSF\ cases, respectively, where $\Gamma _{\text{abs}}^{hh}$
includes the contributions of the $h$ bubble and the bubble of Faddeev-Popov
ghosts, while in the other cases the fields circulating in the loop are
specified by the superscripts.

The contributions $\Gamma _{\text{abs}}^{hh}$ are gauge dependent and can be
collected into field redefinitions, up to cubic corrections (which do not
contribute to $M_{AB}$). Their expressions can be read from formulas
(4.8)-(4.9)\ of ref. \cite{UVQG} in the limit $\alpha \rightarrow 0$, $\xi
\rightarrow 0$. The result is%
\begin{equation}
\Gamma _{\text{abs}}^{hh}=-\int \frac{\delta S_{\text{H}}(g)}{\delta g_{\mu
\nu }}\Delta g_{\mu \nu },\qquad \Delta g_{\mu \nu }=\frac{i\kappa ^{3}}{%
480\pi \zeta }\theta (-\square )\left( 61\square h_{\mu \nu }-42\eta _{\mu
\nu }\square h+42\eta _{\mu \nu }\partial ^{\rho }\partial ^{\sigma }h_{\rho
\sigma }\right) .  \label{form}
\end{equation}%
Formula (\ref{form}) can be rewritten as%
\begin{equation}
\Gamma _{\text{abs}}^{hh}=-\int \frac{\delta S_{\text{QG}}}{\delta g_{\mu
\nu }}\Delta g_{\mu \nu },  \label{equa}
\end{equation}%
up to cubic corrections.

The contributions $\Gamma _{\text{abs}}^{\phi h}$ are also gauge dependent
away from the $\phi $ peak. They can be calculated with the techniques
explained in appendix \ref{appA}. We find 
\begin{equation}
\Gamma _{\text{abs}}^{\phi h}=-\frac{3i\kappa ^{2}m_{\phi }^{2}}{32\pi }\int 
\sqrt{-g}\phi \theta (-\square -m_{\phi }^{2})\left( \square +m_{\phi
}^{2}\right) \frac{1}{\square }\left( 2\square +m_{\phi }^{2}\right) \phi .
\label{form2}
\end{equation}

The other contributions to $\Gamma _{\text{abs}}$ are gauge independent. We
find 
\begin{eqnarray}
\Gamma _{\text{abs}}^{\phi \phi } &=&\frac{i}{16\pi }\frac{1}{120}\int
\left. \sqrt{-g}R_{\mu \nu }\theta (r)\theta (1-r)\sqrt{1-r}\left[
(1-r)^{2}R^{\mu \nu }+\frac{1}{8}(4+12r-r^{2})g^{\mu \nu }R\right]
\right\vert _{g\rightarrow g+2\chi }  \notag \\
&&+\frac{i\kappa }{512\pi }\int \sqrt{-g}\phi \theta (r)\theta (1-r)%
\sqrt{1-r}\left[ 72m_{\phi }^{2}\kappa \phi -r(2+r)\square R\right]
\Big| _{g\rightarrow g+2\chi },  \label{gabsphi}
\end{eqnarray}%
where $r=-4m_{\phi }^{2}/\square $ and the substitutions $g\rightarrow
g+2\chi $ are to be performed on the whole integrands.

Now we turn to $\Gamma _{\text{abs}}^{m}$. Equation (\ref{sgsf}) shows that
these corrections are related to the two-point function of the
energy-momentum tensor $T_{m}^{\mu \nu }$ of equation (\ref{stress}). We can
write 
\begin{equation*}
\Gamma _{\text{abs}}^{m}=\Gamma _{\text{abs}}^{\varphi }+\Gamma _{\text{abs}%
}^{\psi }+\Gamma _{\text{abs}}^{V},
\end{equation*}%
where $\Gamma _{\text{abs}}^{\varphi }$, $\Gamma _{\text{abs}}^{\psi }$ and $%
\Gamma _{\text{abs}}^{V}$ are the contributions of scalar fields, fermions
and gauge vectors, respectively.

On general grounds, the matter fields $\Phi $ of mass $m_{\Phi }$ give an
expression of the form%
\begin{eqnarray}
\Gamma _{\text{abs}}^{\Phi } &=&\frac{i}{16\pi }\int \sqrt{-g}R_{\mu \nu
}\theta (r_{\Phi })\theta (1-r_{\Phi })\sqrt{1-r_{\Phi }}  \notag \\
&&\times \left. \left[ P_{\Phi }(r_{\Phi })\left( R^{\mu \nu }-\frac{1}{3}%
g^{\mu \nu }R\right) +Q_{\Phi }(r_{\Phi })g^{\mu \nu }R\right] \right\vert
_{g\rightarrow (g+2\chi )\mathrm{e}^{\kappa \phi }},  \label{gabsPhi}
\end{eqnarray}%
where $r_{\Phi }=-4m_{\Phi }^{2}/\square $ and $P_{\Phi }(r_{\Phi })$, $%
Q_{\Phi }(r_{\Phi })$ are polynomials that can be calculated as explained in
appendix \ref{appA}.

In the case of $N_{s}$ scalar fields of mass $m_{\varphi }$, with action 
\begin{equation*}
S_{s}=\frac{1}{2}\sum_{i=1}^{N_{s}}\int \sqrt{-g}\left[ g^{\mu \nu
}(\partial _{\mu }\varphi ^{i})(\partial _{\nu }\varphi ^{i})-m_{\varphi
}^{2}\varphi ^{i\hspace{0.01in}2}+\frac{1}{6}(1+2\eta _{s})R\varphi ^{i%
\hspace{0.01in}2}\right] ,
\end{equation*}%
we find that $\Gamma _{\text{abs}}^{\varphi }$ is given by formula (\ref%
{gabsPhi}) with%
\begin{equation}
P_{\varphi }(r)=\frac{N_{s}}{120}(1-r)^{2},\qquad Q_{\varphi }(r)=\frac{N_{s}%
}{576}(4\eta _{s}-r)^{2}.  \label{gscal}
\end{equation}

In the case of $N_{f}$ Dirac fermions of mass $m_{\psi }$, $\Gamma _{\text{%
abs}}^{\psi }$ is given by the same formula with%
\begin{equation*}
P_{\psi }(r)=\frac{N_{f}}{60}\left( 3-r-2r^{2}\right) ,\qquad Q_{\psi }(r)=%
\frac{N_{f}}{144}r(1-r).
\end{equation*}

In the case of $N_{v}$ gauge vectors $V_{\mu }$, $\Gamma _{\text{abs}}^{V}$
is given by%
\begin{equation*}
P_{V}(0)=\frac{N_{v}}{10},\qquad Q_{V}(0)=0.
\end{equation*}

For completeness, we also consider Proca vectors $A_{\mu }$ and Pauli-Fierz
symmetric tensors $\Upsilon _{\mu \nu }$. The Proca action is%
\begin{equation}
S_{\text{P}}(g,A)=\int \sqrt{-g}\left[ -\frac{1}{4}F_{\mu \nu }F^{\mu \nu }+%
\frac{m_{\text{P}}^{2}}{2}A^{\mu }A_{\mu }+\frac{\eta _{\text{P}}}{2}R^{\mu
\nu }A_{\mu }A_{\nu }+\frac{\eta _{\text{P}}^{\prime }}{2}RA^{\mu }A_{\mu }%
\right] ,  \label{ProcaL}
\end{equation}%
where $\eta _{\text{P}}$ and $\eta _{\text{P}}^{\prime }$ parametrize the
nonminimal couplings. The contribution $\Gamma _{\text{abs}}^{\text{P}}$ of $%
N_{\text{P}}$ copies of such vectors to the absorptive part is (\ref{gabsPhi}%
) with%
\begin{equation*}
P_{\text{P}}(r)=\frac{N_{\text{P}}}{120}(13+14r+3r^{2})+P_{\text{P}}^{\text{%
nm}}(r),\qquad Q_{\text{P}}(r)=\frac{N_{\text{P}}}{576}(4-4r+3r^{2})+Q_{%
\text{P}}^{\text{nm}}(r),
\end{equation*}%
where $P_{\text{P}}^{\text{nm}}(r)$ and $Q_{\text{P}}^{\text{nm}}(r)$ are
corrections due to the nonminimal couplings, collected in appendix \ref{appB}%
. A curious fact is that $\Gamma _{\text{abs}}^{\text{P}}$ admits a regular
ultraviolet limit ($m_{\text{P}}\rightarrow 0$) at $\eta _{\text{P}}=\eta _{%
\text{P}}^{\prime }=0$. However, the limit is not conformal invariant, since 
$Q_{\text{P}}(0)\neq 0$, due to simplifications between powers of $m_{\text{P%
}}$ and powers of $1/m_{\text{P}}$. The limit $m_{\text{P}}\rightarrow 0$
does not exist if $\eta _{\text{P}}$ or $\eta _{\text{P}}^{\prime }$ are
nonvanishing.

Equipped with arbitrary nonminimal couplings, parametrized by constants $%
\eta _{i}$, the covariantized action of Pauli-Fierz fields $\Upsilon _{\mu
\nu }$ of mass $m_{\Upsilon }$ reads%
\begin{eqnarray}
\hat{S}_{\text{PF}}(g,\Upsilon ,m_{\Upsilon }^{2}) &=&S_{\text{PF}%
}(g,\Upsilon ,m_{\Upsilon }^{2})+\frac{1}{2}\int \sqrt{-g}\left[ \eta
_{1}R^{\mu \nu \rho \sigma }\Upsilon _{\mu \rho }\Upsilon _{\nu \sigma
}+R^{\mu \nu }(\eta _{2}\Upsilon _{\mu \rho }\Upsilon _{\nu }^{\rho }+\eta
_{3}\Upsilon _{\mu \nu }\Upsilon )\right.  \notag \\
&&\qquad \qquad \left. +R\left( \eta _{4}\Upsilon _{\mu \nu }\Upsilon ^{\mu
\nu }+\eta _{5}\Upsilon ^{2}\right) \right] ,  \label{PFaction}
\end{eqnarray}%
where $\Upsilon =\Upsilon _{\mu \nu }g^{\mu \nu }$. The contribution $\Gamma
_{\text{abs}}^{\text{PF}}$ of such fields to the absorptive part is rather
involved. We report its high-energy behavior in appendix \ref{appB}, enough
to prove that, differently from the case of the Proca fields, no values of
the nonminimal couplings make the ultraviolet limit of $\Gamma _{\text{abs}%
}^{\text{PF}}$ well defined.

In ref. \cite{UVQG} the masses of the matter fields $\Phi $ were set to zero
and both $\phi $ and $\chi _{\mu \nu }$ were implicit and quantized as
fakeons. This means that the results found there apply to the GFF theory at $%
r_{\Phi }=0$. Indeed, it is easy to check that formula (\ref{gabsGFF}) at $%
r_{\Phi }=0$ agrees with the result of \cite{UVQG}, apart from the
expression of $\Delta g_{\mu \nu }$, which is much simpler now. The reason
behind the change of $\Delta g_{\mu \nu }$ is that the two calculations are
done with different classical Lagrangians, (\ref{SQG}) versus (\ref{sew}) or
(\ref{nocosmo}), related by perturbative field redefinitions. Normally, a
change of field variables on the classical action affects the $\Gamma $
functional by modifying the contributions that vanish on the solutions of
the field equations. A general method to work out the change of $\Delta
g_{\mu \nu }$ directly does exist \cite{fieldcov} and requires to extend the
calculations to the composite fields involved in the transformation.

\section{The fakeon width}

\label{probe} \setcounter{equation}{0}

The diagram of Fig. \ref{Fig1} does not allow us to extract physical
quantities for generic values of the center-of-mass energy squared $s=p^{2}$%
, because the graviton gives gauge-dependent contributions, such as (\ref%
{form}) and (\ref{form2}), which\ can be turned into cubic corrections by
means of field redefinitions. If we want a physical quantity for generic $s$%
, the diagrams of Fig. \ref{Fig1} must be accompanied by other diagrams that
contribute to the same order and involve triangles (vertex corrections) and
boxes. Computations of this type have been done extensively in the standard
model \cite{absotriangle} and can be generalized to the theory of quantum
gravity studied here with some effort. However, for the time being, we
concentrate on the fakeon widths, which are physical quantities that can be
extracted just from the bubble diagrams.

Assume that $s$ is very close to $m_{\chi }^{2}$. Then the leading
contributions of the (non-amputated) two-point functions $\langle
G_{A}(p)G_{B}(-p)\rangle $ to $M_{AB}$ are given by $\langle \chi _{\mu \nu
}(p)\chi _{\rho \sigma }(-p)\rangle _{\text{abs}}^{\text{1 loop}}$, which
carry a double pole $1/(s-m_{\chi }^{2})^{2}$. The vertex corrections are
next-to-leading, and so are the contributions such as $\langle h_{\mu \nu
}(p)\chi _{\rho \sigma }(-p)\rangle _{\text{abs}}^{\text{1 loop}}$, since
they give at most simple poles $1/(s-m_{\chi }^{2})$. The gauge-dependent
contributions, such as $\langle h_{\mu \nu }(p)h_{\rho \sigma }(-p)\rangle _{%
\text{abs}}^{\text{1 loop}}$, are next-to-next-to-leading, as are the box
corrections.

This means that the coefficient of the double pole must be physical by
itself at the fakeon peak. For example, it is straightforward to check that
it is gauge independent. Specifically, assuming that the masses of the
matter fields $\Phi $ are much smaller than $m_{\chi }$ and $s\sim m_{\chi }$%
, we find 
\begin{equation}
\langle \chi _{\mu \nu }(p)\chi _{\rho \sigma }(-p)\rangle _{\text{abs}}^{%
\text{1 loop}}=C(s)\frac{\kappa ^{4}}{8\pi \zeta ^{2}}\frac{s^{2}}{%
(s-m_{\chi }^{2})^{2}}\mbox{{\LARGE $\Pi $}}_{\mu \nu \rho \sigma
}^{(2)}(p,s)+\mathcal{O}((s-m_{\chi }^{2})^{0}),  \label{chichiabs}
\end{equation}%
where $\mbox{{\LARGE $\Pi $}}_{\mu \nu \rho \sigma }^{(2)}(p,s)$ can be read
from (\ref{proje}) and%
\begin{equation}
C(s)=C_{m}+C_{\phi }(s),\qquad C_{m}=\frac{N_{s}+6N_{f}+12N_{v}}{120},\qquad
C_{\phi }(s)=\frac{1}{120}\theta (1-r_{\phi })(1-r_{\phi })^{5/2},
\label{ccc}
\end{equation}%
with $r_{\phi }=4m_{\phi }^{2}/s$. Since we are not making assumptions about
the mass of $\phi $, we must take $C_{\phi }(s)$ as a function of $s$.

The quantity $C_{m}$ is known as central charge in conformal field theory.
By analogy, we can define $C(s)$ as the total central charge and $C_{\phi
}(s)$ as the central charge of the massive scalar $\phi $. The function $%
C_{\phi }(s)$ appearing in (\ref{ccc}) holds in the GSF\ theory, where $\phi 
$ is quantized as a physical field, while $C_{\phi }(s)=0$ in the GFF\
theory. The central charges of the graviton and the fakeons are identically
zero.

If we want, we can include $N_{\text{P}}$ Proca vectors with no nonminimal
couplings and small masses $m_{\text{P}}$. Then $C(s)=C_{m}+C_{\text{P}%
}+C_{\phi }(s)$, with 
\begin{equation*}
C_{\text{P}}=\frac{13}{120}N_{\text{P}}.
\end{equation*}%
In the presence of Pauli-Fierz fields $\Upsilon _{\mu \nu }$ and when the
Proca nonminimal couplings are switched on, the total central charge is a
complicated function of $r_{\text{P}}=4m_{\text{P}}^{2}/s$, $r_{\Upsilon
}=4m_{\Upsilon }^{2}/s$ and the nonminimal couplings.

Resumming the self-energies, we can obtain the corrected propagators of the
graviton multiplet. In particular, at the $\chi $ peak we have the two-point
function%
\begin{equation}
\langle \chi _{\mu \nu }(p)\hspace{0.01in}\hspace{0.01in}\chi _{\rho \sigma
}(-p)\rangle _{s\sim \bar{m}_{\chi }^{2}}=-\frac{i\kappa ^{2}}{\zeta }\frac{%
Z_{\chi }}{s-\bar{m}_{\chi }^{2}+i\bar{m}_{\chi }\Gamma _{\chi }}%
\mbox{{\LARGE $\Pi $}}_{\mu \nu \rho \sigma }^{(2)}(p,s),  \label{corrected}
\end{equation}%
where $\bar{m}_{\chi }$ is the corrected $\chi $ mass and $\Gamma _{\chi }$
is the $\chi $ width. We find 
\begin{eqnarray}
\Gamma _{\chi } &=&-\frac{\kappa ^{2}m_{\chi }^{3}}{8\pi \zeta }C+m_{\chi }%
\mathcal{O}\left( \frac{m_{\chi }^{4}}{M_{\text{Pl}}^{4}}\right) =-m_{\chi
}\alpha _{\chi }C+m_{\chi }\mathcal{O}(\alpha _{\chi }^{2}),\qquad  \notag \\
Z_{\chi } &=&1+\mathcal{O}(\alpha _{\chi }),\qquad \bar{m}_{\chi
}^{2}=m_{\chi }^{2}\left[ 1+\mathcal{O}(\alpha _{\chi })\right] ,
\label{nega}
\end{eqnarray}%
where $M_{\text{Pl}}$ is the Planck mass, $C=C(m_{\chi }^{2})$ and $\alpha
_{\chi }=m_{\chi }^{2}/M_{\text{Pl}}^{2}$ is a sort of \textquotedblleft
fakeon/graviton structure constant\textquotedblright .

The negative sign of $\Gamma _{\chi }$ implies that microcausality is
violated. We can illustrate this effect in simple terms by means of the
Breit-Wigner distribution and its Fourier transform. We have%
\begin{equation}
\frac{i}{E-m+i\frac{\Gamma }{2}}\qquad \longrightarrow \qquad \mathrm{sgn}(t)%
\hspace{0.01in}\theta (\Gamma t)\exp \left( -imt-\frac{\Gamma t}{2}\right) ,
\label{breit}
\end{equation}%
so when $\Gamma <0$ the theta function picks the future instead of the past.

Note that the negative overall sign in front of the propagator (\ref%
{corrected}) is consistent with unitarity. Indeed, we find 
\begin{equation}
2\mathrm{Im}\left[ i\langle \chi _{\mu \nu }(p)\hspace{0.01in}\hspace{0.01in}%
\chi _{\rho \sigma }(-p)\rangle _{s\sim \bar{m}_{\chi }^{2}}\right] =-\frac{%
2\kappa ^{2}}{\zeta }\frac{Z_{\chi }\bar{m}_{\chi }\Gamma _{\chi }}{(s-\bar{m%
}_{\chi }^{2})^{2}+\bar{m}_{\chi }^{2}\Gamma _{\chi }^{2}}%
\mbox{{\LARGE $\Pi
$}}_{\mu \nu \rho \sigma }^{(2)}(p,s)\geqslant 0,  \label{unita1}
\end{equation}%
in agreement with the optical theorem. In particular, when we take the limit 
$\Gamma _{\chi }\rightarrow 0^{-}$, we obtain 
\begin{equation}
2\mathrm{Im}\left[ i\langle \chi _{\mu \nu }(p)\hspace{0.01in}\hspace{0.01in}%
\chi _{\rho \sigma }(-p)\rangle _{s\sim \bar{m}_{\chi }^{2}}\right] \underset%
{\Gamma _{\chi }\rightarrow 0^{-}}{\longrightarrow }\frac{2\pi \kappa ^{2}}{%
\zeta }Z_{\chi }\delta (s-\bar{m}_{\chi }^{2})\mbox{{\LARGE $\Pi $}}_{\mu
\nu \rho \sigma }^{(2)}(p,s).  \label{unita2}
\end{equation}

Let us discuss a hypothetical scattering process containing fakeons among
the final states. In that case we must take the imaginary part of the zeroth
order $\chi _{\mu \nu }$ propagator, which however vanishes because of the
quantization prescription (\ref{propchi}). This means that a process of this
type has vanishing cross section and cannot occur. It is impossible to
detect $\chi _{\mu \nu }$ \textquotedblleft before it decays into something
else\textquotedblright , which is consistent with calling $\chi _{\mu \nu }$
a \textquotedblleft fake particle\textquotedblright\ and stating that it
does not belong to the subspace $V$ of the physical fields. The difference
between the peak of a fakeon and the peak of a resonance is that the one of
a fakeon is just a geometric shape and no physical particle is associated
with it. In some respects, this behavior resembles the one of the
\textquotedblleft anomalous thresholds\textquotedblright\ \cite{anomalous}.
In particular, the quantity $1/|\Gamma _{\chi }|$ cannot be viewed as the
lifetime of the fakeon. We could interpret it as the amount of time during
which causality is meaningless. More details are given in the next section,
where we compare the results of the GFF\ and GSF theories with those of the
theory that has ghosts.

If we repeat the calculation around the $\phi $ peak, under the assumption
that the masses of the matter fields are negligible with respect to $m_{\phi
}$, we find the width%
\begin{equation*}
\Gamma _{\phi }=\frac{\eta _{s}^{2}\kappa ^{2}m_{\phi }^{3}}{48\pi \zeta }=%
\frac{m_{\phi }}{6}\alpha _{\phi }\eta _{s}^{2},
\end{equation*}%
where $\alpha _{\phi }\equiv m_{\phi }^{2}/M_{\text{Pl}}^{2}$. We expect
that $\Gamma _{\phi }$ is much smaller than $|\Gamma _{\chi }|$, because it
is only sensitive to the scalar nonminimal coupling $\eta _{s}$. No sign of
microcausality violation is present here, since $\Gamma _{\phi }>0$.

We do not have compelling arguments to predict the values of the masses $%
m_{\chi }$ and $m_{\phi }$, but it is conceivable that they are smaller than
the Planck mass. Taking $m_{\chi }\sim m_{\phi }\sim 10^{11}$GeV, for
definiteness, and assuming the matter content of the standard model ($%
N_{s}=4 $, $N_{f}=45/2$, $N_{v}=12$), we obtain $\alpha _{\chi }\sim 7\cdot
10^{-17}$ and $\Gamma _{\chi }\sim -16$keV. For $m_{\chi }\sim m_{\phi }\sim
10^{12}$GeV we would have $\Gamma _{\chi }\sim -16$MeV. The dumping factor $%
\mathrm{e}^{-\Gamma t/2}$ appearing in formula (\ref{breit}) tells us that
the violation of causality occurs within time intervals of the order of $%
4\cdot 10^{-20}$s for $m_{\chi }\sim 10^{11}$GeV, in the center-of-mass
frame, and $4\cdot 10^{-23}$s for $m_{\chi }\sim 10^{12}$GeV. However, the
oscillating factor $\mathrm{e}^{-imt}$ strongly suppresses those effects up
to energies of the order of the fakeon mass. Other massive particles with
masses smaller than $m_{\chi }$ could be present in nature, besides those
contained in the standard model, and make $C$ and $|\Gamma _{\chi }|$ larger
by one or two orders of magnitude.

More effort is necessary to work out physical quantities away from the
peaks, such as the cross section $\sigma $ as a function of the
center-of-mass energy $\sqrt{s}$. So far we have set the cosmological
constant $\Lambda _{C}$ to zero, but it is not difficult to extend the
calculations to nonvanishing $\Lambda _{C}$.

\section{Comparison with the Stelle theory}

\label{SScompa} \setcounter{equation}{0}

In this section we compare the results found in the GFF and GSF theories
with those that can be obtained in the Stelle GSGh theory, to emphasize the
differences and the effects of the ghosts. The GSGh quantization
prescription is just made of point ($a$) of section \ref{quantization}, so $%
\phi $ is a physical scalar and $\chi _{\mu \nu }$ is a spin-2 ghost.

The absorptive part of the self-energy of the graviton multiplet includes
extra bubble diagrams, whose bubbles are made of: ($i$)\ two $\chi $ legs, ($%
ii$) one $\chi $ leg and one $\phi $ leg, ($iii$) one $\chi $ leg and one $h$
leg. In total, 
\begin{equation*}
\Gamma _{\text{abs}}^{\text{GSGh}}=\Gamma _{\text{abs}}^{\text{GSF}}+\Gamma
_{\text{abs}}^{\chi h}+\Gamma _{\text{abs}}^{\chi \phi }+\Gamma _{\text{abs}%
}^{\chi \chi }.
\end{equation*}%
For the calculations, we use the action (\ref{schi}). The corrections can be
evaluated straightforwardly, but their expressions are quite lengthy, so we
content ourselves with the analysis of the results around the $\chi $ peak.
There, $\Gamma _{\text{abs}}^{\chi \chi }$ does not contribute, because its
threshold is $s=4m_{\chi }^{2}$. Similarly, $\Gamma _{\text{abs}}^{\chi \phi
}$ has a threshold at $s=(m_{\chi }+m_{\phi })^{2}$. Since the $\phi $ mass
is presumably not very different from the $\chi $ mass, $\Gamma _{\text{abs}%
}^{\chi \phi }$ is also negligible at the $\chi $ peak. In the end, only $%
\Gamma _{\text{abs}}^{\chi h}$ is important. We find formula (\ref{chichiabs}%
) with the modified central charge%
\begin{equation}
C(s)=C_{m}+C_{\phi }(s)+C_{\text{Gh}}(s),\qquad C_{\text{Gh}}(s)=-(t_{\chi
}-1)\theta (t_{\chi }-1),  \label{CGh}
\end{equation}%
where $t_{\chi }=s/m_{\chi }^{2}$. The crucial factor $(t_{\chi }-1)$ in $C_{%
\text{Gh}}(s)$, brought by formula (\ref{iabs}), implies that the spin-2
ghost contribution $C_{\text{Gh}}(s)$ to the central charge is subleading
around the peak. The value of $\Gamma _{\chi }$ is still the one of formula (%
\ref{nega}), to the lowest order.

\begin{figure}[t]
\begin{center}
\includegraphics[width=10truecm]{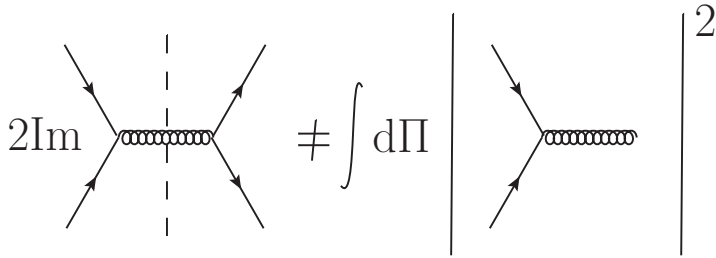}
\end{center}
\caption{Breakdown of the optical theorem in the GSGh theory}
\label{Fig2}
\end{figure}

For sufficiently small time intervals, where the ghost is still
\textquotedblleft alive\textquotedblright , we do not have to resum the
powers of the width $\Gamma _{\chi }$. Indeed, the Stelle theory admits the
process of Fig. \ref{Fig2}, which does not obey the optical theorem. The
right-hand side is positive, while the left-hand side is negative, since%
\begin{eqnarray*}
2\mathrm{Im}\left[ i\langle \chi _{\mu \nu }(p)\hspace{0.01in}\hspace{0.01in}%
\chi _{\rho \sigma }(-p)\rangle _{0}\right] &=&\frac{2\kappa ^{2}}{\zeta }%
\mathrm{Im}\left[ \frac{1}{s-m_{\chi }^{2}+i\epsilon }\right] 
\mbox{{\LARGE
$\Pi $}}_{\mu \nu \rho \sigma }^{(2)}(p,s) \\
&=&-\frac{2\pi \kappa ^{2}}{\zeta }\delta (s-m_{\chi }^{2})%
\mbox{{\LARGE
$\Pi $}}_{\mu \nu \rho \sigma }^{(2)}(p,s)\leqslant 0.
\end{eqnarray*}

The GFF and GSF theories, on the other hand, just give $0=0$ in this case,
since the left-hand side has no imaginary part due to the fakeon
prescription (\ref{nopri}), while the right-hand side vanishes since the
fakeon is projected away from the physical spectrum. This is another way of
saying that the projection in question is consistent.

In some sense, the fakeons can be viewed as \textquotedblleft auxiliary
fields with nontrivial kinetic terms\textquotedblright . They circulate
inside the Feynman diagrams, but cannot enter or exit the diagrams. They
mediate interactions, but cannot be observed directly. We could even
integrate them out (following the rules of the nonanalytic Wick rotation or
the average continuation), but the resulting theory would be nonlocal and
more difficult to handle.

\section{Conclusions}

\label{concl} \setcounter{equation}{0}

We have studied various aspects of the theory of quantum gravity proposed in
ref. \cite{LWgrav}, after converting its higher-derivative action into an
action with two-derivative kinetic terms. The graviton multiplet is made of
the fluctuation $h_{\mu \nu }$ of the metric tensor around flat space, a
massive scalar $\phi $ and a massive spin-2 field $\chi _{\mu \nu }$. The
field $\chi _{\mu \nu }$ is quantized as a fakeon, because its kinetic
action has the wrong overall sign. The scalar $\phi $ can be quantized
either as a fakeon or a physical particle, which leads to two options, the
GFF and GSF theories.

At high energies the nature of $\{h_{\mu \nu },\phi ,\chi _{\mu \nu }\}$ as
a multiplet emerges clearly, the main duty of $\chi _{\mu \nu }$ and $\phi $
being to \textquotedblleft escort\textquotedblright\ the graviton $h_{\mu
\nu }$ and wipe away most ultraviolet divergences it creates, to ensure
renormalizability. At low energies, both $\chi _{\mu \nu }$ and $\phi $
decouple and the ordinary low energy, nonrenormalizable theory is retrieved.

The action of quantum gravity is strictly renormalizable, which makes it
essentially unique (when matter is switched off), because it contains a
finite number of parameters and can be quantized in just two physically
consistent ways. As in the standard model, the matter sector cannot be
predicted from first principles, since it is always possible to add heavy
particles and/or fakeons.

We have calculated the absorptive part of the self-energy of the graviton
multiplet and used it to compute, among other things, the width $\Gamma
_{\chi }$ of the fakeon $\chi _{\mu \nu }$ and the width $\Gamma _{\phi }$
of the scalar $\phi $. The former is negative and proportional to the
central charge $C$. The graviton and the fakeons do not contribute to $C$,
while the other physical fields give positive contributions. Perturbative
unitarity holds, i.e. the optical theorem is satisfied. However, the
negative sign of $\Gamma _{\chi }$ shows that the theory predicts the
violation of causality for center-of-mass energies larger than the fakeon
mass $m_{\chi }$, at distances or time intervals smaller than $1/|\Gamma
_{\chi }|$. There, the notions of past, present and future lose meaning.
Said in different words, the theory implies that causality is not a
principle of nature, but an approximation that is practically useful when
two events are separated by a time interval longer than $1/|\Gamma _{\chi }|$%
. Since at present quantum gravity is the only interaction of nature that
predicts the violation of microcausality, the experimental detection of such
effects could be the first sign that gravity is indeed quantized.

The calculations of this paper can be extended to include the vertex
corrections and the box contributions, along the lines of analogous
computations done in the standard model \cite{absotriangle}, to achieve
gauge independence away from the peaks and obtain the complete cross section 
$\sigma (\sqrt{s})$.

\vskip 10truept \noindent {\large \textbf{Acknowledgments}} \nopagebreak
\vskip 2truept \nopagebreak
We are grateful to U. Aglietti and D. Comelli for useful discussions.

\section*{Appendices}

\renewcommand{\thesection}{A}

\section{Calculations of absorptive parts}

\label{appA}

\setcounter{equation}{0}\renewcommand{\theequation}{\thesection.%
\arabic{equation}}

In this appendix we recall how to calculate the absorptive parts of the
one-loop diagrams. Consider the integral%
\begin{equation*}
I(p,m_{1},m_{2})=\int \frac{\mathrm{d}^{D}k}{(2\pi )^{D}}%
S(k,m_{1})S(p+k,m_{2}),\qquad S(k,m)=\frac{1}{k^{2}-m^{2}+i\epsilon }.
\end{equation*}%
Using the Feynman parameters and renormalizing the divergence away (because
it does not contribute to the absorptive part), we find%
\begin{equation*}
I(p,m_{1},m_{2})=-\frac{i}{16\pi ^{2}}\int_{0}^{1}\mathrm{d}x\ln \left[
m_{1}^{2}x+m_{2}^{2}(1-x)-p^{2}x(1-x)-i\epsilon \right] .
\end{equation*}%
The absorptive part is%
\begin{eqnarray}
I_{\text{abs}}(p,m_{1},m_{2}) &=&-\frac{1}{16\pi }\int_{0}^{1}\mathrm{d}x%
\hspace{0.02in}\theta (p^{2}x(1-x)-m_{1}^{2}x-m_{2}^{2}(1-x))  \notag \\
&=&-\frac{1}{16\pi }\theta \left( p^{2}-(m_{1}+m_{2})^{2}\right) \sqrt{1-%
\frac{(m_{1}+m_{2})^{2}}{p^{2}}}\sqrt{1-\frac{(m_{1}-m_{2})^{2}}{p^{2}}}%
.\qquad  \label{iabs}
\end{eqnarray}%
Similarly, we can treat the \ integrals%
\begin{eqnarray*}
I^{\mu _{1}\cdots \mu _{n}}(p,m_{1},m_{2}) &=&\int \frac{\mathrm{d}^{D}k}{%
(2\pi )^{D}}k^{\mu _{1}}\cdots k^{\mu
_{n}}S(k,m_{1})S(p+k,m_{2})=a_{n}p^{\mu _{1}}\cdots p^{\mu _{n}} \\
&&+a_{n-2}p^{\{\mu _{1}}\cdots p^{\mu _{n-2}}\eta ^{\mu _{n-1}\mu
_{n}\}}+\cdots
\end{eqnarray*}%
(the indices between the curly brackets being completely symmetrized)\ by
expanding the results as sums of polynomials built with $p_{\mu }$ and $\eta
_{\mu \nu }$, multiplied by constants $a_{i}$. The constants are calculated
by contracting with $p_{\mu }$ and $\eta _{\mu \nu }$ and making the
replacements 
\begin{eqnarray*}
k^{2} &\rightarrow &m_{1}^{2},\qquad (p+k)^{2}\rightarrow m_{2}^{2}, \\
p\cdot k &=&\frac{1}{2}\left[ (p+k)^{2}-p^{2}-k^{2}\right] \rightarrow \frac{%
1}{2}\left( m_{2}^{2}-m_{1}^{2}-p^{2}\right)
\end{eqnarray*}%
in the numerators, which follow from the fact that the tadpoles have no
absorptive parts. We get, for example,%
\begin{eqnarray*}
I_{\text{abs}}^{\mu }(p,m_{1},m_{2}) &=&-\frac{p^{\mu }}{2}\left(
1+r_{-}\right) I_{\text{abs}}(p,m_{1},m_{2}), \\
I_{\text{abs}}^{\mu \nu }(p,m_{1},m_{2}) &=&\left[ \frac{p^{\mu }p^{\nu }}{3}%
\left( 1+r_{-}-r_{2}+r_{-}^{2}\right) -\frac{\eta ^{\mu \nu }p^{2}}{12}%
\left( 1-2r_{+}+r_{-}^{2}\right) \right] I_{\text{abs}}(p,m_{1},m_{2}),
\end{eqnarray*}%
where $r_{i}=m_{i}^{2}/p^{2}$ and $r_{\pm }=r_{1}\pm r_{2}$. We can proceed
similarly to work out the expressions of all the $I_{\text{abs}}^{\mu
_{1}\cdots \mu _{n}}(p,m_{1},m_{2})$. For the calculations of this paper, we
just need $n$ from 0 to 4.

\renewcommand{\thesection}{B}

\section{Contributions of Proca and Pauli-Fierz fields}

\label{appB}

\setcounter{equation}{0}\renewcommand{\theequation}{\thesection.%
\arabic{equation}}

Here we collect a few results about the contributions of Proca and
Pauli-Fierz fields to the absorptive part (\ref{gabsPhi}) of the
graviton-multiplet self-energy. The nonminimal couplings of the Proca action
(\ref{ProcaL}) give contributions%
\begin{eqnarray*}
P_{\text{P}}^{\text{nm}}(r) &=&\frac{N_{\text{P}}}{60}\frac{\eta _{\text{P}}%
}{r^{2}}\left[ \eta _{\text{P}}(2+6r+7r^{2})-2r(1+13r+r^{2})\right] , \\
Q_{\text{P}}^{\text{nm}}(r) &=&\frac{N_{\text{P}}}{144r^{2}}\left[ 6\eta _{%
\text{P}}^{\prime }(6\eta _{\text{P}}^{\prime }+r)(4-4r+3r^{2})\right. \\
&&\left. +\eta _{\text{P}}(12\eta _{\text{P}}^{\prime
}+r)(8-10r+5r^{2})+\eta _{\text{P}}^{2}(16-24r+11r^{2})\right] .
\end{eqnarray*}%
The only way to have a smooth ultraviolet limit is by setting $\eta _{\text{P%
}}=\eta _{\text{P}}^{\prime }=0$.

In the case of the Pauli-Fierz action (\ref{PFaction}), we report the first
terms of the high-energy expansion, given by%
\begin{eqnarray*}
P_{\text{PF}}(r) &=&\frac{8N_{\text{PF}}}{135r^{4}}\left[ (3-2\eta _{1}+\eta
_{2})^{2}+\frac{r}{4}(45-6\eta _{1}+5\eta _{2})(3-2\eta _{1}+\eta _{2})+%
\frac{9}{4}r\right] +\mathcal{O}(r^{-2}), \\
Q_{\text{PF}}(r) &=&\frac{4N_{\text{PF}}}{81r^{4}}(3-\eta _{1}+2\eta
_{2}+6\eta _{4})^{2}+\mathcal{O}(r^{-3}).
\end{eqnarray*}%
We see that if we choose the coefficients $\eta _{i}$ of the nonminimal
couplings so that the $\mathcal{O}(r^{-4})$ terms vanish, the $\mathcal{O}%
(r^{-3})$ cannot vanish at the same time. Therefore, it is impossible to
have a smooth ultraviolet limit, in contrast with what happens in the Proca
theory.

\end{document}